\documentclass[aps,twocolumn,pre]{revtex4-1}
\usepackage{graphicx,amsmath,amsfonts,bm, color}

\usepackage{xcolor,hyperref}
\hypersetup{
   colorlinks,
   linkcolor={blue!50!black},
   citecolor={blue!50!black},
   urlcolor={blue!80!black}
}

\usepackage{enumitem}

\renewcommand{\=}{\!=\!}
\newcommand{\1}{^{\mbox{\tiny (1)}}}

\newcommand{\tr}{\operatorname{tr}}

\newcommand{\dbar}{{\,\mathchar'26\mkern-12mu d}}
\DeclareMathAlphabet{\mathitbf}{OML}{cmm}{b}{it}
\newcommand{\rv}{\mathitbf r}

\newcommand{\xv}{\mathitbf x}

\newcommand{\zv}{\mathitbf z}

\newcommand{\calBold}[1]{\mbox{\boldmath${\cal #1}$}}
\def\onedot{$\mathsurround0pt\ldotp$}

\def\cdddot{
  \mathbin{\vcenter{\baselineskip.67ex
    \hbox{\onedot}\hbox{\onedot}\hbox{\onedot}}}}
\begin{document}

\title{Extracting the properties of quasilocalized modes in computer glasses:\\ Long-range continuum fields, contour integrals and boundary effects}
\author{Avraham Moriel$^{1}$}
\author{Yuri Lubomirsky$^{1}$}
\author{Edan Lerner$^{2}$}
\author{Eran Bouchbinder$^{1}$}

\affiliation{$^{1}$Chemical and Biological Physics Department, Weizmann Institute of Science, Rehovot 7610001, Israel\\
$^{2}$Institute for Theoretical Physics, University of Amsterdam, Science Park 904, 1098 XH Amsterdam, The Netherlands}

\begin{abstract}
Low-frequency nonphononic modes and plastic rearrangements in glasses are spatially quasilocalized, i.e.~feature a disorder-induced short-range core and known long-range decaying elastic fields. Extracting the unknown short-range core properties, potentially accessible in computer glasses, is of prime importance. Here we consider a class of contour integrals, performed over the known long-range fields, which are especially designed for extracting the core properties. We first show that in computer glasses of typical sizes used in current studies, the long-range fields of quasilocalized modes experience boundary effects related to the simulation box shape and the widely employed periodic boundary conditions. In particular, image interactions mediated by the box shape and the periodic boundary conditions induce fields' rotation and orientation-dependent suppression of their long-range decay. We then develop a continuum theory that quantitatively predicts these finite-size boundary effects and support it by extensive computer simulations. The theory accounts for the finite-size boundary effects and at the same time allows the extraction of the short-range core properties, such as their typical strain ratios and orientation. The theory is extensively validated in both 2D and 3D. Overall, our results offer a useful tool for extracting the intrinsic core properties of nonphononic modes and plastic rearrangements in computer glasses.
\end{abstract}

\maketitle

\section{Background and motivation}
\label{sec:Intro}

\begin{figure}[ht!]
\centering
\includegraphics[width=0.35\textwidth]{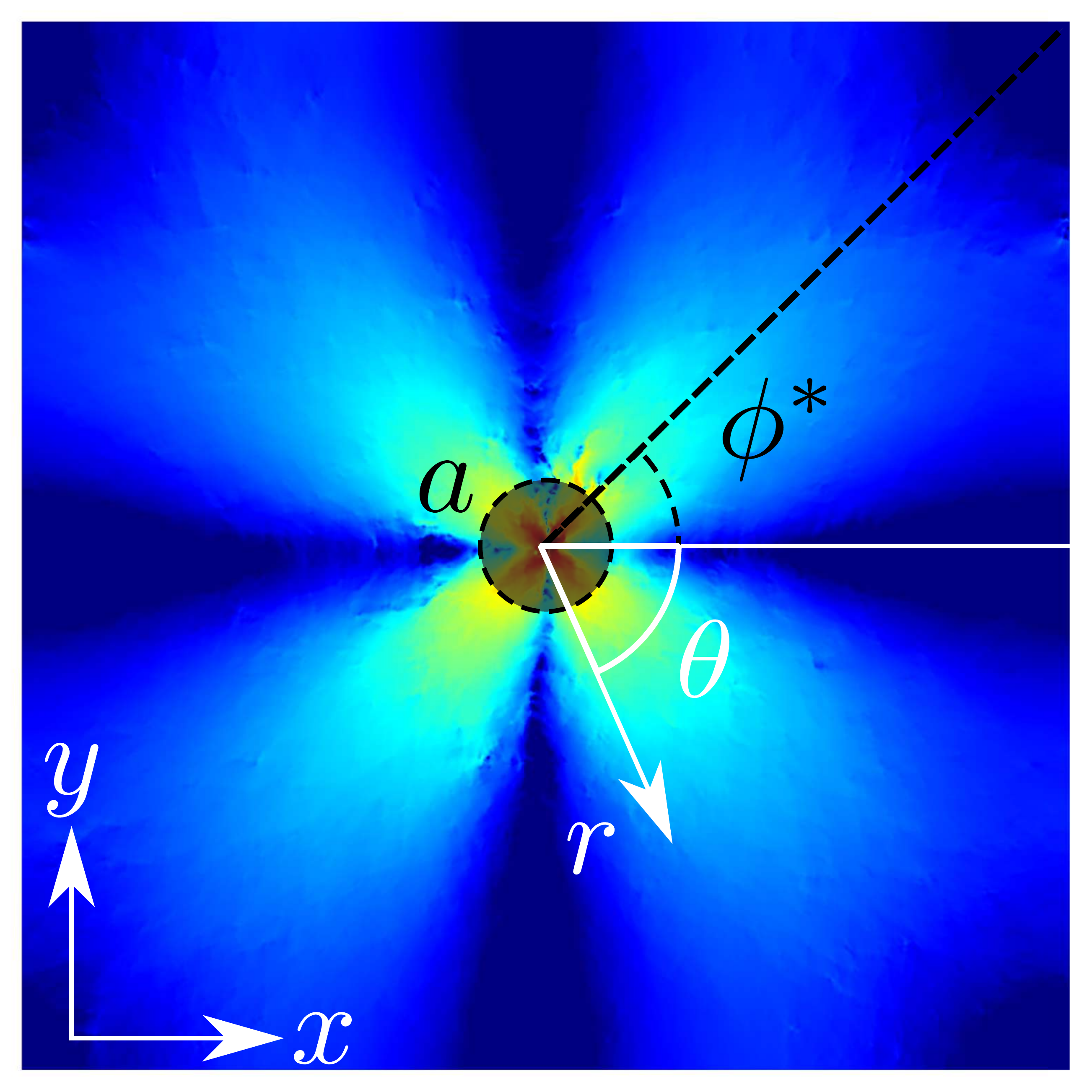}
\caption{An example of a quasilocalized mode in a 2D computer glass. Shown is $\log\!{(|{\bm u}({\bm r})|)}$, the logarithm of the displacement field ${\bm u}({\bm r})$ of a nonlinear quasilocalized mode (see Appendix~\ref{ap:models} for details about the computer glass model and Appendix~\ref{ap:nlmodes} about nonlinear modes). ${\bm r}$ is the position vector relative to the center of the mode (white arrow, represented by the polar coordinates $(r,\theta)$) and hotter/colder colors correspond to larger/smaller displacements. The mode exhibits intense displacements at its core (marked by a dashed circle) of linear size $a$, which are accompanied by a long-range decaying field. The mode also exhibits azimuthal quadrupolar structure~\cite{lemaitre2004,lemaitre2006_avalanches} oriented at an angle $\phi^*$ relative to a fixed Cartesian coordinate system ($x,y$), which is aligned with the simulation box (bottom left corner).}
\label{fig:fig1}
\end{figure}

Structural disorder in glassy materials gives rise to physical phenomena absent from their ordered crystalline counterparts. A notable example is the emergence of quasilocalized modes, either in the form of low-frequency nonphononic excitations in the absence of external driving forces~\cite{soft_potential_model_1991,Gurevich2003, modes_prl_2016,lte_pnas,ikeda_pnas, modes_prl_2018,atsushi_core_size_pre,cge_paper,LB_modes_2019,modes_prl_2020,pinching_pnas,corrado_kappa_statistics_jcp_2020} or in the form of quasilocalized irreversible (plastic) rearrangements under external driving forces~\cite{spaepen_1977,argon_st,falk_langer_stz,lemaitre2004,Argon_prb_2005,micromechanics2016}. Quasilocalized modes feature a short-range disordered core and long-range decaying displacement fields. The latter follow a power-law $\sim\!1/r^{\dbar-1}$ \cite{lemaitre2006_avalanches,modes_prl_2016,atsushi_core_size_pre} for $r\!\gg\!a$, where $r$ is the distance from the center of the mode, $a$ is the linear size of the core and $\dbar$ is the spatial dimension. An example of such a mode in $\dbar\=2$ is presented in Fig.~\ref{fig:fig1}, see figure caption for details.

The statistical-mechanical properties of quasilocalized modes significantly affect the thermodynamic~\cite{Zeller_and_Pohl_prb_1971,Anderson,Phillips,soft_potential_model_1991,ultrastable_perez_pnas}, transport~\cite{Buchenau_prb_1992_soft_potential_model_transport, Ikeda_scattering_2018, scattering_jcp}, and strongly nonlinear and dissipative properties of glassy materials~\cite{argon_st,argon_bubble_raft, argon_simulations, dennin_foam_stzs, pine_emulsions_stzs, schall_stz_colloids, falk_langer_stz, lemaitre2004, lemaitre2006_avalanches, Vasoya2016, micromechanics2016, MW_prx_mean_field_yielding, MW_yielding_2018_pre, Ozawa6656}. Consequently, elucidating their spatial and geometric properties is an important step in understanding the physics of glasses. While much is known about the scaling properties of quasilocalized modes' long-range fields \cite{lemaitre2006_avalanches,modes_prl_2016}, far less is known about the properties of their short-range cores, emerging from microstructural disorder \cite{Rodney_2016,Rodney_2017,rosso_qlms_soft_matter_2018,Rottler2018,atsushi_core_size_pre,pinching_pnas}. In particular, the strain (displacement gradients) amplitudes inside the core, the orientation of the mode (cf.~Fig.~\ref{fig:fig1}), the statistical distributions of these quantities and their dependence on the glass history and driving forces are not yet fully characterized. As the core size $a$ is microscopic in nature, typically of the order of a few atomic lengths, the short-range core properties are inaccessible in laboratory molecular glasses. As a result, computer simulations of model glasses play a central role in exploring the physics of quasilocalized modes~\cite{Schober_Laird_numerics_PRL,SchoberOligschleger1996,falk_langer_stz, lemaitre2004,widmer2008irreversible,modes_prl_2016, lte_pnas, ikeda_pnas, modes_prl_2018,atsushi_core_size_pre,cge_paper,LB_modes_2019,modes_prl_2020,pinching_pnas,corrado_kappa_statistics_jcp_2020}. Yet, to the best of our knowledge, we still lack systematic, robust and efficient approaches for extracting the short-range core properties in computer glasses. The main goal of this paper is to develop and substantiate such an approach.

Several recent works pursued a similar goal~\cite{rosso_qlms_soft_matter_2018,Rodney2016,Rodney2017,Rottler2018}. The approach developed in this paper bears some resemblance to various aspects of these recent works, but also differs from them quite significantly, both in its premises and outcomes --- we highlight both the similarities and the differences below. In what follows, we propose and test an approach for extracting the short-range core properties of quasilocalized modes in computer glasses based on the long-range fields, and in particular on a set of contour integrals that are designed to reveal the short-range core properties.

In Sect.~\ref{sec:contour}, we discuss the proposed set of contour integrals based on the long-range continuum fields obtained for infinite media, under the assumption that proper scale separation is achieved in computer glasses of typical sizes used in current studies. We demonstrate that in some cases the contour integrals allow the extraction of the short-range properties, while in others this approach fails. In Sect.~\ref{sec:rotation}, we show that the deviations from the infinite medium theory are related to the core orientation, and demonstrate orientation-dependent fields' rotation and the suppression of their long-range decay. In Sect.~\ref{sec:image}, we show that these observations are related to image interactions due to the periodic boundary conditions commonly employed. We develop a continuum theory of image interactions and their boundary effects in finite-size computer glasses, and show that it quantitatively explains in a unified manner the observed deviations from the infinite medium predictions. The resulting formalism then allows extracting short-range core properties in computer glasses of typical sizes. In Sect.~\ref{sec:validation}, we extensively validate the continuum-derived measures in both 2D and 3D against an independent microscopic measure of the core orientation and by a direct comparison to the atomistic quasilocalized modes in computer glasses. Finally, in Sect.~\ref{sec:summary} we offer some concluding remarks.

\section{Extracting short-range core properties using the long-range continuum fields}
\label{sec:contour}

The existence of the long-range fields of quasilocalized modes in glasses is a direct consequence of the localized deformation that defines the short-range core. Hence, the former encodes information about the latter, and our goal here is to develop a formalism that allows the extraction of the core properties from the long-range fields alone. This physical situation is similar in nature to other known examples, e.g.~dislocations in crystalline materials~\cite{hirth1983theory}. There, the long-range fields encode information about the magnitude and orientation of the Burgers vector, which quantifies the topological defect that characterizes the dislocation core~\cite{Landau_Lifshitz_Elasticity}. The dislocation core properties can be extracted by performing closed-path contour integration over the long-range fields. While nonphononic excitations and irreversible (plastic) rearrangements in glassy materials are not topological line defects like dislocations in ordered crystalline materials, a similar approach can nevertheless be developed for them as well.

To see this, we first note that this general class of problems can be addressed using Eshelby's inclusions formalism~\cite{Eshelby,Eshelby2}. In this formalism, the core of linear size $a$ (i.e.~the inclusion) is assumed to undergo a homogeneous inelastic deformation characterized by the so-called eigenstrain tensor $\bm{\mathcal{E}}^*$ (which is not diagonal). The main result relevant for our purposes here is that the displacement vector field ${\bm u}({\bm r})$ outside the core/inclusion (${\bm r}$ is the position vector relative to the center of the core/inclusion, cf.~Fig.~\ref{fig:fig1}) is expressed as an integral over the core volume, $u_i(\bm{r})\=-C_{jklm}\,\mathcal{E}^*_{lm}\!\int_{v}\partial_{k} G_{ij}(\bm{r}-\bm{r'})d\bm{r'}$. Here $\bm{C}$ is the elastic stiffness tensor, indices represent Cartesian components and $v\!\propto\!a^\dbar$ is the $\dbar$-dimensional core/inclusion volume.
${\bm G}({\bm r})$ is the linear elastic Green's function of infinite isotropic media, whose Fourier transform reads~\cite{wilmanski2010}
\begin{equation}
\label{eq:Green_FT}
  \bm{G}(\bm{q}) = \frac{1}{\mu}\left[\frac{\bm{\mathcal{I}}_\dbar}{q^2} - \frac{\lambda + \mu}{\lambda + 2\mu}\frac{\bm{q}\otimes\bm{q}}{q^4}\right] \ ,
\end{equation}
where $\bm{q}$ is the $\dbar$-dimensional wave vector, $\bm{\mathcal{I}}_{\dbar}$ is the $\dbar$-dimensional identity tensor and $\otimes$ is a diadic product. Focusing on the far-field, $r\!\gg\!a$, $-\!\int_{v}\partial_{k} G_{ij}(\bm{r}-\bm{r'})d\bm{r'}$ is well approximated by $-v\partial_k G_{ij}(\bm{r})$, leading to
\begin{equation}
\label{eq:disp_lr}
  u_i(\bm{r})\simeq -v\,C_{jklm}\,\mathcal{E}^*_{lm}\,\partial_k G_{ij}(\bm{r}) \ .
\end{equation}
Note that $\bm{C}$ is assumed here to be spatially homogeneous and that for isotropic media it can be fully expressed in terms of the Lam\'e constants $\lambda$ and $\mu$, or, equivalently, in terms of the shear and bulk moduli~\cite{Landau_Lifshitz_Elasticity}.

The core strain tensor $\bm{\mathcal{E}}^*$, like any other second-rank tensor, can be split into its dilatational (isotropic) part, $\bm{\mathcal{E}}^*_{\mbox{\scriptsize dil}} \!=\!\tfrac{1}{\dbar}\tr\!\left(\bm{\mathcal{E}}^*\right)\bm{\mathcal{I}}_{\dbar}\!=\!\epsilon_{\mbox{\scriptsize dil}}^*\,\bm{\mathcal{I}}_{\dbar}$ ($\epsilon_{\mbox{\scriptsize dil}}^*$ is the dilatational eigenstrain), and its deviatoric part, $\bm{\mathcal{E}}^*_{\mbox{\scriptsize dev}} \!=\!\bm{\mathcal{E}}^*\! -  \bm{\mathcal{E}}^*_{\mbox{\scriptsize dil}}$. The deviatoric part may be decomposed as $\bm{\mathcal{E}}^*_{\mbox{\scriptsize dev}}\!=\!\bm{P}(\bm{\phi^*})\,\bm{\epsilon}_{\mbox{\scriptsize dev}}^*\,\bm{P}^T\!(\bm{\phi^*})$, i.e.~as a rotation $\bm{P}(\bm{\phi^*})$ of the diagonal deviatoric core tensor $\bm{\epsilon}_{\mbox{\scriptsize dev}}^*$ by the generalized angles $\bm{\phi}^*$. As $\bm{\mathcal{E}}^*_{\mbox{\scriptsize dev}}$ is real and symmetric, $\bm{P}(\bm{\phi^*})$ is a real orthogonal matrix, $\bm{P}^{-1}(\bm{\phi^*})\!=\!\bm{P}^T(\bm{\phi^*})$, depending on $\dbar(\dbar\!-\!1)/2$ generalized angles $\bm{\phi^*}$. The diagonal deviatoric tensor $\bm{\epsilon}_{\mbox{\scriptsize dev}}^*$, which satisfies $\tr(\bm{\epsilon}_{\mbox{\scriptsize dev}}^*)\=0$, contains $\dbar\!-\!1$ independent strain amplitudes. Together with the generalized angles, which determine the orientation of the core, the deviatoric part of $\bm{\mathcal{E}}^*$ is characterized by $(\dbar - 1)(2+\dbar)/2$ independent numbers, while the dilatational part is characterized by a single number (the dilatational eigenstrain $\epsilon_{\mbox{\scriptsize dil}}^*$). Our goal is to use Eq.~\eqref{eq:disp_lr}, assuming ${\bm u}(\bm{r})$ is known or measured far from the core ($r\!\gg\!a$), in order to extract these independent numbers.

To see how all this works, we first specialize to 2D infinite media, i.e.~set $\dbar\=2$ (the 3D case is addressed below in Subsect.~\ref{subsec:3D}) and do not consider boundary effects. Taking the 2D inverse Fourier transform of $\bm{G}(\bm{q})$ in Eq.~\eqref{eq:Green_FT}, one obtains
\begin{equation}
\label{eq:G_2D}
\bm{G}(\bm{r})\!=\!\frac{\lambda + \mu}{4 \pi\mu \left(\lambda + 2\mu\right)}\left[\frac{\rv\!\otimes\!\rv}{r^2} - \frac{\lambda + 3\mu}{\lambda + \mu}\log(r)\,\bm{\mathcal{I}}_2\right] \ ,
\end{equation}
where $r\=|{\bm r}|$. Moreover, the core strain tensor $\bm{\mathcal{E}}^*$ in 2D can be expressed as
\begin{equation}
\label{eq:core2D}
  \bm{\mathcal{E}}^* = \bm{P}(\phi^*) \,\bm{\epsilon}_{\mbox{\scriptsize dev}}^*\,\bm{P}(-\phi^*) + \epsilon_{\mbox{\scriptsize dil}}^*\,\bm{\mathcal{I}}_2 \ ,
\end{equation}
where $\bm{P}(\phi^*)\!=\!\Bigl(\!\begin{smallmatrix}
  \cos(\phi^*) & {\tiny -}\!\sin(\phi^*) \\
   \sin(\phi^*) & \,\cos(\phi^*)
\end{smallmatrix}\!\Bigr)$, $\bm{\epsilon}_{\mbox{\scriptsize dev}}^*\!=\!\text{diag}\left(\epsilon^*_{\mbox{\scriptsize dev}},-\epsilon^*_{\mbox{\scriptsize dev}}\right)$ (characterized by a single deviatoric strain amplitude $\epsilon^*_{\mbox{\scriptsize dev}}$), $\phi^*$ is the orientation of the core (cf.~Fig.~\ref{fig:fig1}) and $\epsilon_{\mbox{\scriptsize dil}}^*$ is the dilatational eigenstrain.

We next define on the left-hand-sides of Eqs.~\eqref{eq:I_0}-\eqref{eq:I_2S} a set of closed-path contour (azimuthal) integrals over the displacement field ${\bm u}({\bm r})$. We then use the 2D $G_{ij}(\bm{r})$ and $\mathcal{E}^*_{ij}$ of the previous paragraph inside Eq.~\eqref{eq:disp_lr}, together with $C_{ijkl}$ for homogeneous and isotropic media (expressed in terms of $\lambda$ and $\mu$~\cite{Landau_Lifshitz_Elasticity}), to obtain ${\bm u}({\bm r})$ in the large $r$ limit ($r\!\gg\!a$). Evaluating the contour integrals for the resulting ${\bm u}({\bm r})$, we obtain the following $r$-independent limits on the right-hand-sides
\begin{subequations}
\label{eq:Int_2D}
\begin{align}
  I_0(r) &\equiv \frac{1}{2} \left(\frac{\lambda + 2 \mu}{\lambda + \mu}\right) \int_{0}^{2\pi} \bm{u}(\bm{r})\!\cdot\!\bm{r}\,d\theta \xrightarrow[]{r\gg a}  v\epsilon_{\mbox{\scriptsize dil}}^* \ , \label{eq:I_0}\\
  I^{{\scriptscriptstyle (1)}}_{2}(r) &\equiv  \int_{0}^{2\pi} \bm{u}(\bm{r})\!\cdot\!\bm{r}\,\cos(2\theta)\,d\theta \xrightarrow[]{r\gg a} v\epsilon_{\mbox{\scriptsize dev}}^* \cos(2\phi^*) \ , \label{eq:I_2C} \\
  I^{{\scriptscriptstyle (2)}}_{2}(r) &\equiv  \int_{0}^{2\pi} \bm{u}(\bm{r})\!\cdot\!\bm{r}\,\sin(2\theta)\,d\theta \xrightarrow[]{r\gg a}  v\epsilon_{\mbox{\scriptsize dev}}^* \sin(2\phi^*) \ . \label{eq:I_2S}
\end{align}
\end{subequations}
Here we used polar coordinates $(r,\theta)$ to represent the position vector ${\bm r}$ and note that the azimuthal angle $\theta$ should not be confused with the core orientation $\phi^*$ (cf.~Fig.~\ref{fig:fig1}). Equations~\eqref{eq:I_0}-\eqref{eq:I_2S} show, as is also evident from Eq.~\eqref{eq:disp_lr}, that the core area $v\!\propto\!a^2$ cannot be disentangled from the strain amplitudes, and only $v\epsilon_{\mbox{\scriptsize dil}}^*$, $\bar{I}_2(r)\!\equiv\!\sqrt{[I^{{\scriptscriptstyle (1)}}_{2}(r)]^2 \!+\! [I^{{\scriptscriptstyle (2)}}_{2}(r)]^2}\!\xrightarrow[]{r\gg a}\!v\epsilon_{\mbox{\scriptsize dev}}^*$ and $\tfrac{1}{2}\arctan\!\left[I^{{\scriptscriptstyle (2)}}_{2}(r)/I^{{\scriptscriptstyle (1)}}_{2}(r)\right]\!\xrightarrow[]{r\gg a}\!\phi^*$ can be extracted using this approach.

To the best of our knowledge, the set of integrals in Eqs.~\eqref{eq:I_0}-\eqref{eq:I_2S} has not been proposed before in the literature, even though recent works \cite{Rodney2016,Rodney2017,Rottler2018} employed  Eshelby's out-of-inclusion fields for similar purposes. The Eshelby's fields based approach developed in~\cite{Rodney2016} differs from ours in two major respects; first, it is based on a brute force fitting of the 3D Eshelby's fields to the numerical displacements (in fact, multiple quaslilocalized modes have been fitted simultaneously). Second, it was applied to the full-field solution, including the near-field ($r\!\simeq\!a$) part, i.e.~not focusing on the large $r$ limit (the far-field, $r\!\gg\!a$) as we do here. A similar fitting procedure to the full-field Eshelby 2D solution has been employed earlier in~\cite{Carmel2013} in order to extract the short-range core properties.

In~\cite{Rottler2018}, the focus was on extracting the orientation of the core in 2D, i.e.~$\phi^*$. To that aim, a method based on azimuthal Fourier decomposition has been proposed and tested, in addition to employing the fitting procedure of~\cite{Rodney2017}. The azimuthal Fourier modes approach~\cite{Rottler2018} has not been applied directly to the atomistic displacement field ${\bm u}(\bm{r})$, but rather to a related coarse-grained strain field.

Our next goal is to test the validity and utility of the predictions in Eqs.~\eqref{eq:I_0}-\eqref{eq:I_2S}, using the long-range part of $\bm{u}(\bm{r})$ of quasilocalized modes in computer glasses. To that aim, one should first consider several pertinent issues. First, Eqs.~\eqref{eq:I_0}-\eqref{eq:I_2S} are expected to be valid in the large $r$ limit, $r\!\gg\!a$, and therefore the linear size of the simulation box $L$ of the computer glass should be properly selected so as to resolve this limit. As $a$ is estimated to equal a few atomic lengths (i.e.~a few particle sizes $a_0$ in simulations) this should not pose a serious constraint and choosing $L\!\simeq\!50a$, for example, seems to be sufficient. In particular, for such linear system sizes one expects that for $a\!\ll\!r\!\ll\!L$ the integrals on the left-hand-side of Eqs.~\eqref{eq:I_0}-\eqref{eq:I_2S} would feature $r$-independent plateaus and that finite-size effects related to the widely employed periodic boundary conditions would appear at $r\!\lesssim\!L$.

Another relevant issue is the selection of isolated quasilocalized modes to be tested and their identification in computer glasses. Harmonic (linear) nonphononic excitations in the absence of external driving forces, i.e.~quasilocalized normal modes of a glass at zero temperature~\cite{Schober_Laird_numerics_PRL,modes_prl_2016}, are not easily identified due to their prevalent hybridization with extended phononic excitations~\cite{SciPost2016,phonon_widths,episode_1_2020}. Plastic rearrangements, on the other hand, are decoupled from extended phononic excitations under external driving forces; yet, they are not easily identified at finite temperatures (due to thermal fluctuations) and typically lead to additional rearrangements in the limit of zero temperature, resulting in multiple coexisting quasilocalized modes (plastic avalanches) \cite{lemaitre2004_avalanches, lemaitre2006_avalanches, salerno_robbins, Barrat_inertia}.
\begin{figure}[ht!]
\centering
\includegraphics[width=0.48\textwidth]{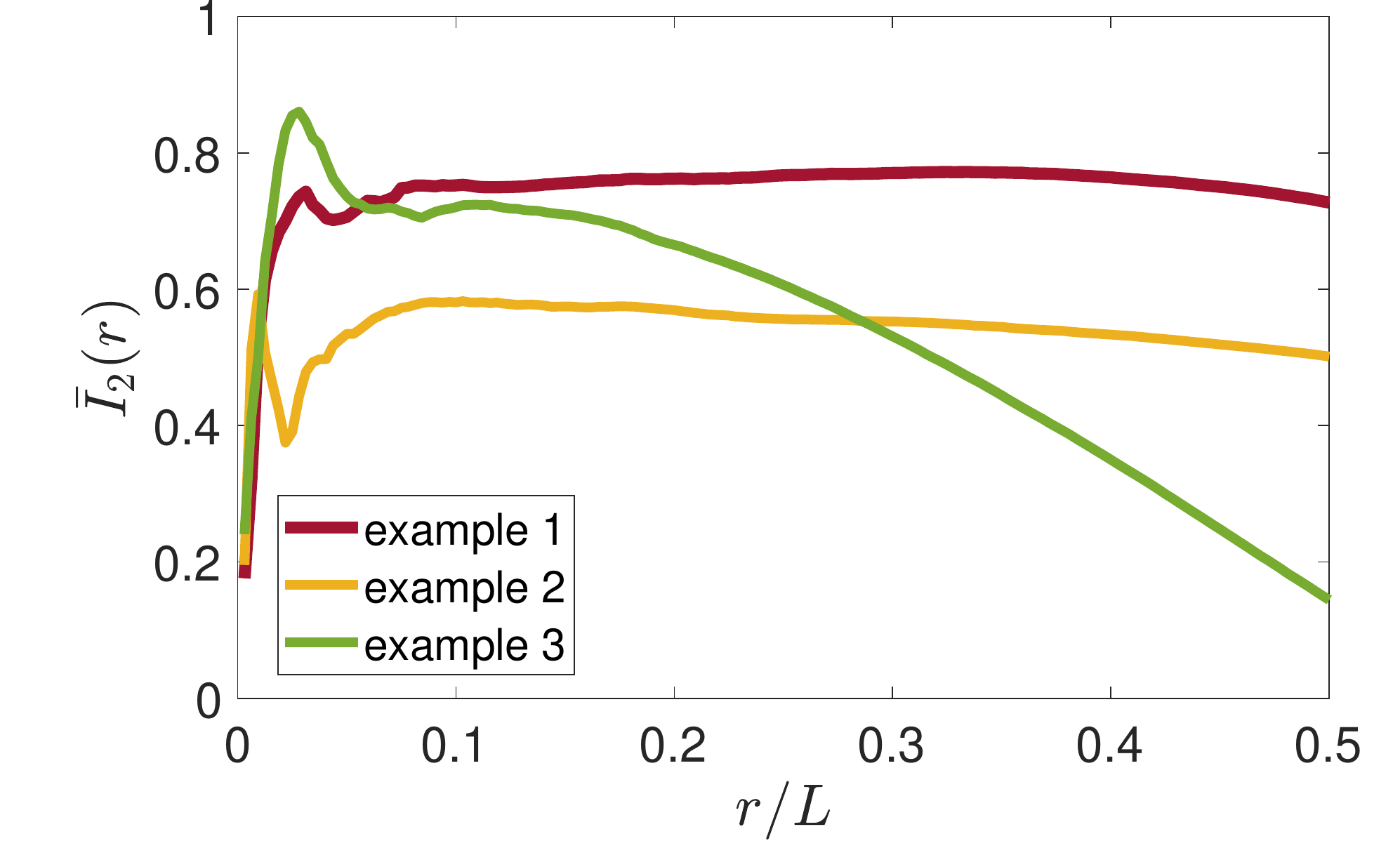}
\caption{$\bar{I}_2(r)\!=\!\sqrt{[I^{{\scriptscriptstyle (1)}}_{2}(r)]^2\!+\![I^{{\scriptscriptstyle (2)}}_{2}(r)]^2}$ vs.~$r/L$, cf.~Eqs.~\eqref{eq:I_2C}-\eqref{eq:I_2S}, for three different nonlinear quasilocalized modes ${\bm u}({\bm r})$ identified in a 2D computer glass with $L\!=\!345a_0$ (see Appendix~\ref{ap:models} for additional information about the computer glass model, and Appendix~\ref{ap:nlmodes} for information about how the nonlinear modes were identified and calculated). The presented results are discussed in detail in the text.}
\label{fig:fig2}
\end{figure}

\begin{figure*}[ht!]
\centering
\includegraphics[width=\textwidth]{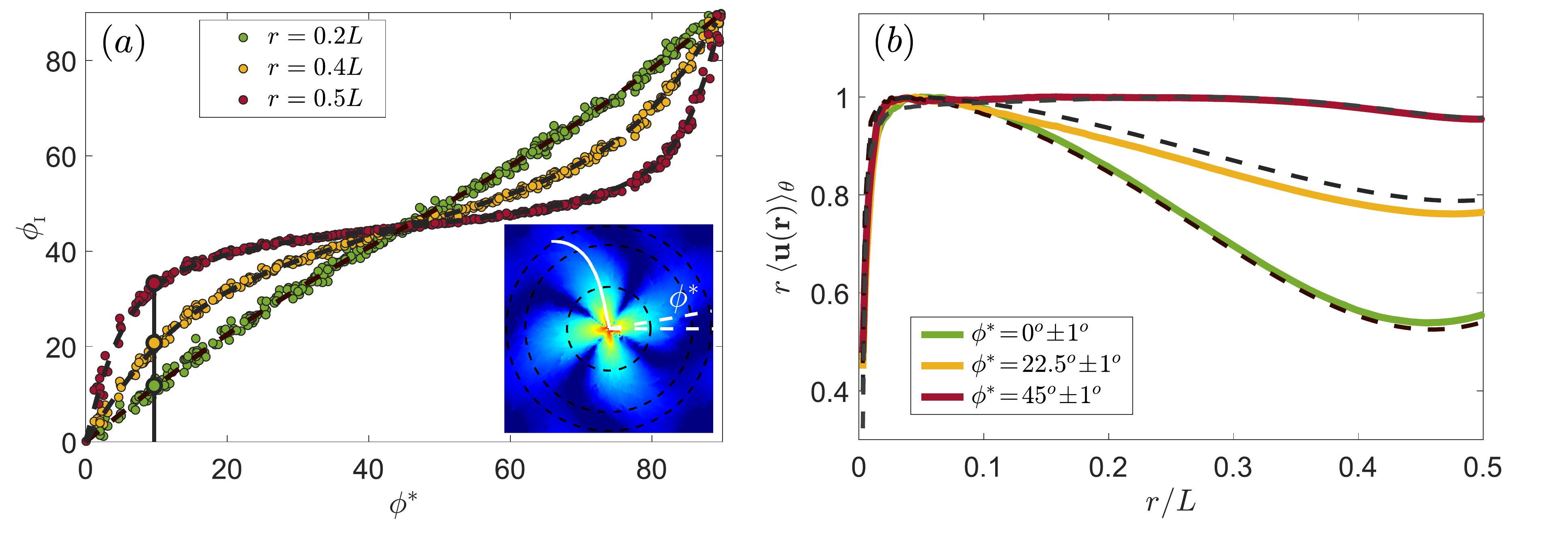}
\caption{(a) $\phi_{\mbox{\tiny I}}(r)\!\equiv\!\tfrac{1}{2}\arctan\!\left[I^{{\scriptscriptstyle (2)}}_{2}(r)/I^{{\scriptscriptstyle (1)}}_{2}(r)\right]$, where $I^{{\scriptscriptstyle (1)}}_{2}(r)$ and $I^{{\scriptscriptstyle (2)}}_{2}(r)$ are defined in Eqs.~\eqref{eq:I_2C}-\eqref{eq:I_2S}, vs.~$\phi^*$ (obtained using the microscopic measure of Sect.~\ref{subsec:microscopic}) for many nonlinear quasiocalized  modes and $r\!=\!0.2L, 0.4L, 0.5L$ (discrete symbols, see legend). The results indicate orientation-dependent fields' rotation --- see text for additional details, explanations and discussion ---, as is explicitly demonstrated in the inset. The inset presents a nonlinear quasilocalized  mode with $\phi^*\!=\!9.7^\circ$ (cf.~Fig.~\ref{fig:fig1}), which corresponds to the vertical line in the main panel. The white line shows $\phi_{\mbox{\tiny I}}(r)$, rotated by $90^\circ$ relative to $\phi^*$ for visual clarity. The dashed line circles correspond to $r\!=\!0.2L, 0.4L, 0.5L$, i.e.~to the $3$ intersections of the vertical line with the data points in the main panel (marked by thick circles). The dashed lines in the main panel are the theoretical predictions of the image interaction theory of Sect.~\ref{sec:image}, see text for details. (b) $r\,\langle{\bm u}({\bm r})\rangle_\theta$ (solid lines) averaged over many atomistic quasilocalized modes with $\phi^*\!=\!0^\circ\!\pm\!1^\circ, 22.5^\circ\!\pm\!1^\circ, 45^\circ\!\pm\!1^\circ$ (see legend), as a function of $r/L$ ($\langle{\bm u}({\bm r})\rangle_\theta\!\equiv\!\sqrt{\!\int_0^{2\pi\!}|{\bm u}({\bm r})|^2\,d\theta}$). The theoretical predictions of the image interaction theory of Sect.~\ref{sec:image} are superimposed (dashed lines). Both the atomistic and theoretical curves are normalized by their maximal value. See text for additional details, explanations and discussion.}
\label{fig:fig3}
\end{figure*}
To address these issues, we choose here to analyze nonlinear nonphononic excitations, which are a family of quasilocalized modes that generalize quasilocalized harmonic (linear) normal modes~\cite{SciPost2016,micromechanics2016,episode_1_2020}. Beyond their general importance for glass physics~\cite{cge_paper,episode_1_2020}, they are particularly useful for our purposes here because they are cleanly identified as they do not hybridize with phononic excitations in the absence of external forces, because they can be identified one at a time and because they are representatives of plastic rearrangements~\cite{micromechanics2016,phm_letter_2020}. In particular, nonlinear modes share the same long-range fields with other quasilocalized modes in glasses~\cite{SciPost2016} and hence are suitable for testing the suggested approach. We stress that the approach developed in this paper can be equally applied to other quasilocalized modes in glasses, for example plastic rearrangements.

In Fig.~\ref{fig:fig2} we present examples of $\bar{I}_2(r)\=\sqrt{[I^{{\scriptscriptstyle (1)}}_{2}(r)]^2\!+\![I^{{\scriptscriptstyle (2)}}_{2}(r)]^2}$, cf.~Eqs.~\eqref{eq:I_2C}-\eqref{eq:I_2S}, for three different nonlinear quasilocalized modes ${\bm u}({\bm r})$ identified in a 2D computer glass with $L\=345a_0$ (see Appendix~\ref{ap:models} for additional information about the computer glass model, and Appendix~\ref{ap:nlmodes} for information about how the nonlinear modes were identified and calculated). In all three examples, $\bar{I}_2(r)$ increases with $r$ at short distances, which we identify with the core of quasilocalized modes. Beyond a certain distance, which indicates the core size $a$, the $\bar{I}_2(r)$ curves appear to reach a plateau level. In the first two examples this plateau level persists over large distances, which we identify with the range $a\!\ll\!r\!\ll\!L$, until $\bar{I}_2(r)$ slightly decreases for $r\!\simeq\!L$. This is exactly the behavior predicted by Eqs.~\eqref{eq:I_2C}-\eqref{eq:I_2S} and hence the robust plateau level can be identified with $v\epsilon_{\mbox{\scriptsize dev}}^*$. On the other hand, the plateau in the third example is very short and subsequently $\bar{I}_2(r)$ significantly decreases with increasing $r$, in sharp contrast with the predictions of Eqs.~\eqref{eq:I_2C}-\eqref{eq:I_2S}.

Compiling a large set of examples in a large ensemble of computer glass realizations, cf.~Appendix~\ref{ap:models} for details, we confirmed that the picture emerging from Fig.~\ref{fig:fig2} is representative. That is, for many nonlinear quasilocalized modes, the predictions of Eqs.~\eqref{eq:I_0}-\eqref{eq:I_2S} are perfectly satisfied and the short-range core properties can be robustly extracted using the proposed approach, while for others the predictions seem to badly fail. Our goal in the next two sections is to understand these rather puzzling observations and to extend the proposed approach to allow the extraction of the short-range core properties under all circumstances.

\section{The core orientation and long-range fields' rotation}
\label{sec:rotation}

What is the physical origin of the failure of the theoretical predictions in Eqs.~\eqref{eq:I_0}-\eqref{eq:I_2S} in some cases? What distinguishes the cases in which they seem to be valid from those in which they fail? To start addressing these questions, we try first to gain additional insight regarding the latter. To that aim, we consider the quantity $\tfrac{1}{2}\arctan\!\left[I^{{\scriptscriptstyle (2)}}_{2}(r)/I^{{\scriptscriptstyle (1)}}_{2}(r)\right]$, where $I^{{\scriptscriptstyle (1)}}_{2}(r)$ and $I^{{\scriptscriptstyle (2)}}_{2}(r)$ are the integrals defined in Eqs.~\eqref{eq:I_2C}-\eqref{eq:I_2S}. Sufficiently away from the core, i.e.~for $r$ sufficiently larger than $a$, we expect this quantity to correspond to the core orientation $\phi^*$. That is, if we define $\phi_{\mbox{\tiny I}}(r)\!\equiv\!\tfrac{1}{2}\arctan\!\left[I^{{\scriptscriptstyle (2)}}_{2}(r)/I^{{\scriptscriptstyle (1)}}_{2}(r)\right]$, we expect $\phi_{\mbox{\tiny I}}(r)\!\xrightarrow[]{r\gg a}\!\phi^*$, as was already stated below Eqs.~\eqref{eq:I_0}-\eqref{eq:I_2S}.

To test this prediction, we need to calculate $\phi_{\mbox{\tiny I}}(r)$ for a large number of quasilocalized modes and different $r$'s, and compare it to an independent measure of the core orientation $\phi^*$. Such an independent measure of $\phi^*$, based on completely different considerations, is discussed in Sect.~\ref{subsec:microscopic}. For our purposes here, we just need to accept the statement that another approach can reliably extract the core orientation $\phi^*$ of any given quasilocalized mode. Accepting it, we applied this approach to quasilocalized nonlinear modes (see Sect.~\ref{subsec:microscopic} for details), obtaining $\phi^*$ for each of them. For each mode ${\bm u}({\bm r})$, we also calculated $\phi_{\mbox{\tiny I}}(r)$ for $r\=0.2L, 0.4L, 0.5L$. These three values of $r$ have been selected because all of them seem to satisfy $r\!\gg\!a$ (here $L\=345a_0$ as in Fig.~\ref{fig:fig2}).

In Fig.~\ref{fig:fig3}a, we plot (discrete symbols) $\phi_{\mbox{\tiny I}}(r)$ vs.~$\phi^*$ for many nonlinear quasiocalized  modes and $r\=0.2L, 0.4L, 0.5L$. For $r\=0.2L$, we observe that all of the data points lie on a straight line of unity slope and no intercept, i.e.~the prediction $\phi_{\mbox{\tiny I}}(r)\!\xrightarrow[]{r\gg a}\!\phi^*$ is satisfied. This, however, is not the case for $r\=0.4L$ and $r\=0.5L$, where deviations from the prediction are observed, except for modes with $\phi^*\!\approx\!45^{\circ}$. In light of these observations, we plot in the inset of Fig.~\ref{fig:fig3}a a quasilocalized nonlinear mode with $\phi^*\!\approx\!10^{\circ}$. It is explicitly observed that for $r\=0.2L$ (inner circle) the mode is oriented at the core angle $\phi^*$, while for $r\=0.4L$ and $r\=0.5L$ (two outer circles) it exhibits systematic deviations from $\phi^*$ (see also the vertical line at $\phi^*\!\approx\!10^{\circ}$ in the main panel). Therefore, depending on the core orientation $\phi^*$, quasilocalized modes in our computer simulations feature long-range fields' rotation. Note that the mode presented in Fig.~\ref{fig:fig1}, which has $\phi^*\!\approx\!45^{\circ}$, does not feature such a long-range fields' rotation, consistently with Fig.~\ref{fig:fig3}a (main panel).

The results presented in Fig.~\ref{fig:fig3}a therefore raise the hypothesis that what distinguishes the cases in which the theoretical predictions in Eqs.~\eqref{eq:I_0}-\eqref{eq:I_2S} are valid from those in which they fail (cf.~Fig.~\ref{fig:fig2}) is the core orientation $\phi^*$. A quick check of the three examples presented in Fig.~\ref{fig:fig2} reveals that the two modes that correspond to the long plateaus (which agree with the theoretical prediction) feature $\phi^*\!\approx\!45^{\circ}$, while the third one, which exhibits a significantly shorter plateau, features a significantly different orientation. With this insight in mind, we performed the integrals in Eqs.~\eqref{eq:I_2C}-\eqref{eq:I_2S} for a large number of quasilocalized nonlinear modes and classified the results according to the core orientation $\phi^*$ of each mode. As we are interested in the spatial decay of $\bar{I}_2(r)\=\sqrt{[I^{{\scriptscriptstyle (1)}}_{2}(r)]^2\!+\![I^{{\scriptscriptstyle (2)}}_{2}(r)]^2}$, whose integrand is proportional to $\bm{u}(\bm{r})\!\cdot\!\bm{r}$, we focused on $r\,\langle{\bm u}({\bm r})\rangle_\theta$, where $\langle{\bm u}({\bm r})\rangle_\theta\!\equiv\!\sqrt{\!\int_0^{2\pi\!}|{\bm u}({\bm r})|^2\,d\theta}$.

In Fig.~\ref{fig:fig3}b (solid lines), we present $r\,\langle{\bm u}({\bm r})\rangle_\theta$ averaged over many quasilocalized modes with $\phi^*\=0^\circ\!\pm\!1^\circ, 22.5^\circ\!\pm\!1^\circ, 45^\circ\!\pm\!1^\circ$, as a function of $r$. It is observed that modes with $\phi^*\!\approx\!45^\circ$ feature a long plateau, which implies that ${\bm u}({\bm r})$ for such modes decays as $1/r$ over a significant fraction of the simulation box, as predicted by the infinite medium theory for $\dbar\=2$. The curves for the other $\phi^*$ values significantly deviate from the predicted plateau, indicting orientation-dependent suppression of the predicted long-range fields. These results are similar to those presented in Fig.~\ref{fig:fig2}, collectively showing that the orientation-dependent suppression of the predicted long-range fields and the orientation-dependent long-range fields' rotation are intrinsically related. Our next goal is to understand these observations in a unified theoretical manner.

\section{Continuum theory of image interactions and boundary effects in finite-size computer glasses}
\label{sec:image}

In order to address the orientation-dependent fields' rotation and the suppression of their long-range decay discussed in the previous two sections, we need to revisit the assumptions behind Eqs.~\eqref{eq:I_0}-\eqref{eq:I_2S} and reassess whether they are satisfied in the computer simulations. The formal assumption behind Eqs.~\eqref{eq:I_0}-\eqref{eq:I_2S} is that ${\bm u}({\bm r})$ is dominated by the long-range power-law fields $\sim\!1/r^{\dbar-1}$. This, in turn, is expected be realized far from the short-range core of an isolated quasilocalized mode (i.e.~one that does not interact with other modes) in a large enough system.

Computer glass simulations are commonly performed under periodic boundary conditions with an elementary simulation box of linear size $L$~\cite{allen1989computer}. Under such conditions, even if there exists a single quasilocalized mode in the elementary simulation box, this mode interacts with its images in the other copies of the elementary (original) box through the periodic boundary conditions. Taking $L$ to be sufficiently large, we expect these image interactions to be sufficiently weak in the spatial range $a\!\ll\!r\!\ll\!L$, where the long-range power-law fields $\sim\!1/r^{\dbar-1}$ are expected to be realized. Naively, taking $L\!=\!345a_0$ as in the examples of Fig.~\ref{fig:fig2}, which is about $50$ times the core size $a$, should be enough.

To quantitatively predict the box size $L$ needed in order to properly resolve the long-range power-law fields $\sim\!1/r^{\dbar-1}$, one needs to calculate the finite-size corrections to the theoretical results presented in Sect.~\ref{sec:contour} due to the periodic boundary conditions. To that aim, we first derive the finite-size periodic boundary conditions counterpart of the infinite medium Green's function in Eq.~\eqref{eq:G_2D}. This is simply achieved by calculating the inverse Fourier series of $\bm{G}(\bm{q})$ in Eq.~\eqref{eq:Green_FT} over the discrete set of Fourier ${\bm q}$-modes allowed by the periodic boundary conditions, obtaining $\bm{G}^{\circ}(\bm{r})\= \sum_{\bm{q}\in\bm{Q}}e^{2 \pi i \bm{q} \cdot \bm{r}}\,\bm{G}(\bm{q})$ in any dimension. Here the $\circ$ denotes periodic boundary conditions and $\bm{Q}$ denotes the range of allowed values of $\bm{q}$, e.g.~$\left(q_x,q_y\right)\!\in\!\tfrac{1}{L}\left(n,m\right)$ with $\left(n,m\right)\!\in\!\mathbb{Z}^2$ in 2D ($\dbar\=2$). Finally, as we are still interested in the spatial range $r\!\gg\!a$, Eq.~\eqref{eq:disp_lr} remains valid, and the displacement field $\bm{u}^{\circ}(\bm{r})$ is obtained by plugging into it $\bm{G}^{\circ}(\bm{r})$ instead of $\bm{G}(\bm{r})$.

With $\bm{u}^{\circ}(\bm{r})$ at hand, we can now test whether the finite-size periodic boundary conditions theory, which takes into account image interactions, quantitatively accounts for the available observations. To that aim, we first generate synthetic quasilocalized modes $\bm{u}^{\circ}(\bm{r})$ using the image interactions theory with various core orientations $\phi^*$, $\epsilon_{\mbox{\scriptsize dil}}^*\=0$ and an arbitrary fixed $\epsilon_{\mbox{\scriptsize dev}}^*$, cf.~Eq.~\eqref{eq:core2D} and the inline equations below it, for $L$ as in Figs.~\ref{fig:fig2} and~\ref{fig:fig3}. We then calculate $\phi_{\mbox{\tiny I}}(r)\=\tfrac{1}{2}\arctan\!\left[I^{{\scriptscriptstyle (2)}}_{2}(r)/I^{{\scriptscriptstyle (1)}}_{2}(r)\right]$ using $\bm{u}^{\circ}(\bm{r})$ inside Eqs.~\eqref{eq:I_2C}-\eqref{eq:I_2S} for $r\=0.2L, 0.4L, 0.5L$, and superimpose the (theoretical) results (dashed lines) on top of the numerical ones in Fig.~\ref{fig:fig3}a. The theoretical results perfectly agree with the numerical ones, providing strong evidence that the origin of orientation-dependent fields' rotation observed in our computer simulations is indeed image interactions induced by the periodic boundary conditions imposed on the finite simulation box. Note that for $\phi^*\!\approx\!45^\circ$ (cf.~Fig.~\ref{fig:fig3}a), the symmetry of the mode and that of the simulation box agree, i.e.~the mode is aligned with the diagonal of the box, and hence image interactions do not lead to rotation.

The very same continuum theory is also expected to account for the orientation-dependent suppression of the long-range fields predicted by the infinite medium theory. To test this, we use $\bm{u}^{\circ}(\bm{r})$ as above for $\phi^*\=0^\circ, 22.5^\circ, 45^\circ$, and calculated $r\,\langle{\bm u}^{\circ}({\bm r})\rangle_\theta\=r\,\sqrt{\!\int_0^{2\pi\!}|{\bm u}^{\circ}({\bm r})|^2\,d\theta}$. The (theoretical) results (dashed lines) are then superimposed on top of the numerical ones in Fig~\ref{fig:fig3}b. It is again observed that the image interactions theory nicely predicts the atomistic data. We therefore conclude that despite the original naive expectation, the selected $L$ in our simulations was not large enough to properly resolve the $1/r^{\dbar-1}$ fields under all circumstances, i.e.~for all core orientations $\phi^*$. We note in passing that image interactions have been claimed not to play a dominant role in the 3D simulations of~\cite{Rodney2016,Rodney2017} and they have not been discussed at all in~\cite{Carmel2013, Rottler2018}.

It is important to stress that the image interaction picture emerging from Fig.~\ref{fig:fig2} and Fig.~\ref{fig:fig3}, and from the theory that explains it, remains valid independently of the value of $L$, as long as periodic boundary conditions are employed and when considering the rescaled spatial variable $r/L$. Yet, the behavior of the contour integrals in Eqs.~\eqref{eq:I_0}-\eqref{eq:I_2S} does depend on $L$ when considered as a function of $r/a$. In particular, increasing $L$ will result in an extended $a\!\ll\!r\!\ll\!L$ spatial region and hence will indeed allow better resolving the $1/r^{\dbar-1}$ fields of quasilocalized modes with any core orientation $\phi^*$.

In order to make progress in relation to the main goal of this paper, i.e.~extracting the core properties of quasilocalized modes in computer glasses, we need to make a pragmatic decision at this stage, in light of the available results. One possibility is to perform simulations with significantly larger $L$'s such that the infinite medium predictions of Eqs.~\eqref{eq:I_0}-\eqref{eq:I_2S} are properly resolved for all orientations. This possibility involves a non-negligible computational cost. Alternatively, as the image interactions tend to suppress the long-range fields at a distance from the core comparable to $L$ (cf.~Figs.~\ref{fig:fig2} and \ref{fig:fig3}a), one can estimate the core properties on the right-hand-sides of Eqs.~\eqref{eq:I_0}-\eqref{eq:I_2S} at the position in which the largest contour integral attains its {\em maximal value}. This maximal value is expected to occur on the plateau of the contour integral, when image interactions are weak, or is expected to probe the prediction of the infinite medium theory, when image interactions are strong. In the next section, this suggestion is extensively tested and validated.

\section{Testing and validating the continuum approach in 2D and 3D}
\label{sec:validation}

Our goal in this section is to test the theoretical framework developed above. To this aim, we first provide in Subsect.~\ref{subsec:3D} the details of the theory in 3D. Next, in Subsect.~\ref{subsec:microscopic} we develop a microscopic measure that independently extracts the core orientation, which is then compared to the continuum measure's predictions in 3D (the corresponding 2D comparison has already been presented in Fig.~\ref{fig:fig3}a). Finally, in Subsect.~\ref{subsec:full-field} we present a direct comparison between atomistic quasilocalized modes and the corresponding continuum framework in 2D and 3D. Overall, the presented results strongly support the developed continuum tool for extracting the core properties of quasilocalized modes in computer glasses.

\subsection{The 3D continuum approach}
\label{subsec:3D}

The continuum theory developed in Sects.~\ref{sec:contour} and~\ref{sec:image} is general, i.e.~dimension-independent. Yet, fully explicit expressions and examples have been provided only in 2D so far. Here we provide explicit expressions also in 3D, where examples follow. The starting point is the Fourier transform of the Green's function in Eq.~\eqref{eq:Green_FT}, whose inverse transform in 3D reads~\cite{wilmanski2010,kachanov2013}
\begin{equation}
\label{eq:Green_3D}
  \bm{G}(\bm{r}) = \frac{\lambda + \mu}{8 \pi\mu \left(\lambda + 2 \mu\right)}\left[\frac{\rv\!\otimes\!\rv}{r^3} + \frac{\lambda + 3\mu}{\lambda + \mu}\frac{\bm{\mathcal{I}}_3}{r} \right] \ .
\end{equation}
Using then Eq.~\eqref{eq:Green_3D}, together with Eq.~\eqref{eq:disp_lr}, we construct the following set of surface integrals
\begin{subequations}
\begin{align}
 & I_0(r)\!\equiv\! 2\sqrt{\pi}\!\left(\!\frac{\!\lambda + 2 \mu\!}{3 \lambda + 2 \mu}\!\right)\!\!\int_{S}\!\bm{u}(\bm{r})\!\cdot\!\bm{r}\,Y_0^0(\Omega)\,r\,d\Omega \xrightarrow[]{r\gg a}  v\epsilon_{\mbox{\scriptsize dil}}^*, \label{eq:3DI_0} \\
 & I_2^{{\scriptscriptstyle(m)}}(r) \equiv  2\sqrt{5\pi} \left(\frac{\lambda + 2 \mu}{3 \lambda + 5 \mu}\right) \!\int_{S}\!\bm{u}(\bm{r})\!\cdot\!\bm{r}\,Y_2^m(\Omega)\,r\,d\Omega \ , \label{eq:3DI_m}
  \end{align}
\end{subequations}
where the surface integral $S$ is performed on a sphere of radius $r$, $Y_2^m(\Omega)$ are the real (i.e.~not complex) orthogonal spherical harmonics of the second degree and order $m\=-2,-1,0,1,2$~\cite{Blanco1997,chisholm1976} (see Eq.~(6) in~\cite{Blanco1997}), and $\Omega$ is the solid angle.

Equations~\eqref{eq:3DI_0}-\eqref{eq:3DI_m} are the 3D counterparts of the 2D Eqs.~\eqref{eq:I_0}-\eqref{eq:I_2S}. In Eq.~\eqref{eq:3DI_m} (which in fact represents $5$ different equations, corresponding to $m\=-2,-1,0,1,2$), unlike Eqs.~\eqref{eq:I_2C}-\eqref{eq:I_2S}, we do not provide explicit expressions in the $r\!\gg\!a$ limit, simply because these are too lengthy. The latter depend on $5$ independent quantities: $3$ generalized angles $\bm{\phi}^*$ that we quantify below through the Euler angles $(\phi^*,\varphi^*,\psi^*)$ (instead of $1$ in 2D) and $2$ deviatoric eigenstrains (multiplied by the core volume $v$) denoted by $v\epsilon_{\mbox{\scriptsize dev,1}}^*$ and $v\epsilon_{\mbox{\scriptsize dev,2}}^*$ (instead of $1$ in 2D). The third one is given by $v\epsilon_{\mbox{\scriptsize dev,3}}^*\=-(v\epsilon_{\mbox{\scriptsize dev,1}}^*+v\epsilon_{\mbox{\scriptsize dev,2}}^*)$.

To extract these $5$ core properties in 3D from the $r\!\gg\!a$ limit of the integrals $I_2^{{\scriptscriptstyle(m)}}(r)$, we first construct the tensor
\begin{equation}
\label{eq:matrix}
  \bar{\bm{I}}_2(r) = \left(\begin{array}{ccc}
                   -\frac{1}{2}I_{2}^{{\scriptscriptstyle(0)}}\!+\!\frac{\sqrt{3}}{2} I_{2}^{{\scriptscriptstyle (2)}} & \frac{\sqrt{3}}{2} I_{2}^{{\scriptscriptstyle(-2)}} & \frac{\sqrt{3}}{2} I_{2}^{{\scriptscriptstyle(1)}} \\
                   \frac{\sqrt{3}}{2} I_{2}^{{\scriptscriptstyle(-2)}} & -\frac{1}{2}I_{2}^{{\scriptscriptstyle(0)}}\!-\!\frac{\sqrt{3}}{2} I_{2}^{{\scriptscriptstyle (2)}} & \frac{\sqrt{3}}{2} I_{2}^{{\scriptscriptstyle(-1)}} \\
                   \frac{\sqrt{3}}{2} I_{2}^{{\scriptscriptstyle(1)}} & \frac{\sqrt{3}}{2} I_{2}^{{\scriptscriptstyle(-1)}} &  I_{2}^{{\scriptscriptstyle(0)}}
                 \end{array}\right) \ ,
\end{equation}
following~\cite{Blanco1997} (cf.~Table 1 therein). Using a few simple test cases, we verified that the eigenvalues of $\bar{\bm{I}}_2(r)$ in the $r\!\gg\!a$ limit, denoted by $\bar{I}_{2}^{{\scriptscriptstyle(i)}}$ (with $i\=1\!-\!3$), satisfy $\bar{I}_{2}^{{\scriptscriptstyle(i)}}\=v\epsilon_{\mbox{\scriptsize dev,i}}^*$ and that the principal directions of the diagonalizing rotation matrix $\bm{P}(\bm{\phi^*})$ (see definition in Sect.~\ref{sec:contour}) correspond to the core Euler angles $(\phi^*,\varphi^*,\psi^*)$ (note that in Subsect.~\ref{subsec:microscopic} we also use the notation $(\phi_{\mbox{\tiny I}}, \varphi_{\mbox{\tiny I}}, \psi_{\mbox{\tiny I}})$, when this approach is compared to the results of an independent approach). Consequently, diagonalizing $\bar{\bm{I}}_2(r)$ of Eq.~\eqref{eq:matrix} allows --- in principle --- to extract the core properties in 3D. Finally, to apply the image interaction theory of Sect.~\ref{sec:image}, we again use $\bm{G}^{\circ}(\bm{r})\=\sum_{\bm{q}\in\bm{Q}}e^{2 \pi i \bm{q} \cdot \bm{r}}\,\bm{G}(\bm{q})$, but this time $\bm{Q}$ corresponds to $\left(q_x,q_y,q_z\right)\!\in\!\tfrac{1}{L}\left(n,m,l\right)$ with $\left(n,m,l\right)\!\in\!\mathbb{Z}^3$, and $\bm{u}^{\circ}(\bm{r})$ is obtained by plugging $\bm{G}^{\circ}(\bm{r})$ into Eq.~\eqref{eq:disp_lr} (instead of $\bm{G}(\bm{r})$). The core properties are evaluated at the distance $r$ where the largest $|\bar{I}_{2}^{{\scriptscriptstyle(i)}}(r)|$ attains its maximum, as will be further detailed below.

\subsection{A microscopic measure of the core orientation and its comparison to the continuum measure}
\label{subsec:microscopic}

In order to test the continuum approach developed above, we propose here an alternative/complementary approach for extracting the core orientation. It is a microscopic approach that makes no reference to the long-range continuum fields, but rather relies on the intrinsic anisotropic structure of quasilocalized modes. This approach has already been used in Fig.~\ref{fig:fig3}a in comparison to the 2D continuum approach, and our goal here is to define it in detail and use it also to independently test the continuum approach in 3D.

A natural way to probe the orientational structure of quasilocalized modes is to look at the way they couple to an external field of a well-defined orientation, in particular to an applied strain tensor $\bm{\epsilon}$. In order to quantify this coupling, we first define a simple scalar characterizer of the displacement field $\bm{u}$ of quasilocalized modes, i.e.~its energy/stiffness $\kappa(\bm{u})\!\equiv\!\hat{\bm{u}}\cdot\frac{\partial^2U(\bm{x})}{\partial\xv\partial\xv}\cdot\hat{\bm{u}} = \hat{u}_i\frac{\partial^2U(\bm{x})}{\partial x_i \partial x_j}\hat{u}_j$ (Einstein's summation convention is assumed). Here $\hat{\bm{u}}\!\equiv\!\bm{u}/|\bm{u}|$ is a $\dbar N$-dimensional unit vector pointing in the direction of $\bm{u}$ ($\xv$ denotes the \emph{particles'} coordinates, to be distinguished from the coordinate vector $\rv$ used in the continuum approach above) and $U(\bm{x})$ is the potential energy of the system.

The coupling between $\bm{u}$ and $\bm{\epsilon}$ can be then quantified through the derivative $d\kappa/d\bm{\epsilon}$. An explicit expression for $d\kappa/d\bm{\epsilon}$ is obtained in the framework of the micro-mechanical theory of nonlinear quasilocalized modes~\cite{micromechanics2016}. In particular, the nonlinear quasilocalized modes used for the analysis above --- the so-called cubic nonlinear modes $\hat{\bm{\pi}}$~\cite{SciPost2016,episode_1_2020}, were shown to satisfy~\cite{micromechanics2016}
\begin{equation}
\label{eq:kappa_eom}
\frac{d\kappa}{d\bm{\epsilon}} \simeq -\frac{\tau}{\kappa} \left(\frac{\partial^2U}{\partial\bm{\epsilon}\partial\xv}\cdot \hat{\bm{\pi}}\right) \ ,
\end{equation}
where $\tau\!\equiv\!\frac{\partial U}{\partial \bm{x}\partial \bm{x}\partial \bm{x}}\cdddot\hat{\bm{\pi}}\hat{\bm{\pi}}\hat{\bm{\pi}}=\frac{\partial U}{\partial x_i \partial x_j\partial x_k}\hat{\pi}_i\hat{\pi}_j\hat{\pi}_k$ (here $\cdddot$ is a triple contraction and Einstein's summation convention is used), and the contraction in $\frac{\partial^2U}{\partial\bm{\epsilon}\partial\xv}\!\cdot\!\hat{\bm{\pi}}$ is understood to be performed over the spatial coordinates $\xv$ and the mode's spatial components.

Equation~\eqref{eq:kappa_eom} determines the change of $\kappa$ with a general applied strain $\bm{\epsilon}$, taking the form $d\kappa/d\bm{\epsilon}\!\sim\!\kappa^{-1}$. The strength of the coupling between $\bm{u}$ and $\bm{\epsilon}$ is encapsulated in the magnitude of the prefactor; when the applied strain is aligned with the mode, this prefactor is expected to be large, while in other directions it is expected to be significantly smaller. Therefore, the prefactor is expected to encode the orientational information we are interested in. Finally, as $\tau$ is independent of the external strain tensor $\bm{\epsilon}$, the coupling to $\bm{\epsilon}$ is contained in the last term on the right-hand-side of Eq.~\eqref{eq:kappa_eom}. Generalizing this coupling term to any quasilocalized mode $\bm{u}$, we define the tensor
\begin{equation}
\label{eq:calF}
\calBold{F} \equiv -\frac{\partial^2U}{\partial\bm{\epsilon}\partial\xv} \cdot \hat{\bm{u}} \ .
\end{equation}
To construct a scalar coupling strength out of the tensor $\calBold{F}$, and in order to relate it to a spatial orientation, we consider a unit vector in $\dbar$ spatial dimensions $\hat{\bm{e}}$ and the squared magnitude of its projection on $\calBold{F}$, i.e.~ $\|\calBold{F}\cdot\hat{\bm{e}}\|^2\!\equiv\!\hat{\bm{e}}^T\!\cdot\!\calBold{F}^T\!\cdot\!\calBold{F}\!\cdot\!\hat{\bm{e}}$.

$\|\calBold{F}\cdot\hat{\bm{e}}\|^2$ depends on the relative orientation of $\hat{\bm{e}}$ and the core orientation $\hat{\bm{e}}^*$ (in 2D, as in Sect.~\ref{sec:rotation}, the latter is defined by a single angle $\phi^*$. In 3D, three angles --- e.g.~the Euler angles --- are required). A basic theorem in linear algebra~\cite{strang1993} states that $\|\calBold{F}\cdot\hat{\bm{e}}\|^2\!\le\!\lambda_{\mbox{\scriptsize max}}^2$, where $\lambda_{\mbox{\scriptsize max}}$ is the largest eigenvalue (in absolute value) of the symmetric tensor $\calBold{F}$. Furthermore, the orientation $\hat{\bm{e}}$ for which $\|\calBold{F}\cdot\hat{\bm{e}}\|^2$ is maximal, i.e.~$\|\calBold{F}\cdot\hat{\bm{e}}\|^2\=\lambda_{\mbox{\scriptsize max}}^2$, is the eigenvector corresponding to $\lambda{\mbox{\scriptsize max}}$~\cite{strang1993}. This orientation is nothing but the core orientation $\hat{\bm{e}}^*$, i.e.~the core orientation $\hat{\bm{e}}^*$ corresponds to the maximum of $\|\calBold{F}\cdot\hat{\bm{e}}\|^2$  with respect to all possible orientations $\hat{\bm{e}}$.

Pragmatically, the core orientation $\hat{\bm{e}}^*$ is simply obtained by diagonalizing $\calBold{F}$ and finding the eigenvector corresponding to its largest eigenvalue. This procedure has been used in Sect.~\ref{sec:image} to extract the core orientation $\phi^*$ in 2D and to compare it in Fig.~\ref{fig:fig3}a to the corresponding continuum measure of the orientation, $\phi_{\mbox{\tiny I}}$. This comparison revealed excellent quantitative agreement between the two approaches, lending strong support to both of them. In order to extract the Euler angles $(\phi^*, \varphi^*, \psi^*)$, a single orientation vector $\hat{\bm{e}}^*$ is not sufficient, rather the other eigenvectors should be obtained as well. We note in passing that the proposed microscopic approach for finding the orientation of quasilocalized modes, when applied to the 2D case, can be related to Eq.~(2) in~\cite{xu2019atomic}, which has been proposed in a different context. Moreover, a different 2D microscopic approach has been discussed in~\cite{Rottler2018}.

\begin{figure}[ht]
\centering
\includegraphics[width=0.4\textwidth]{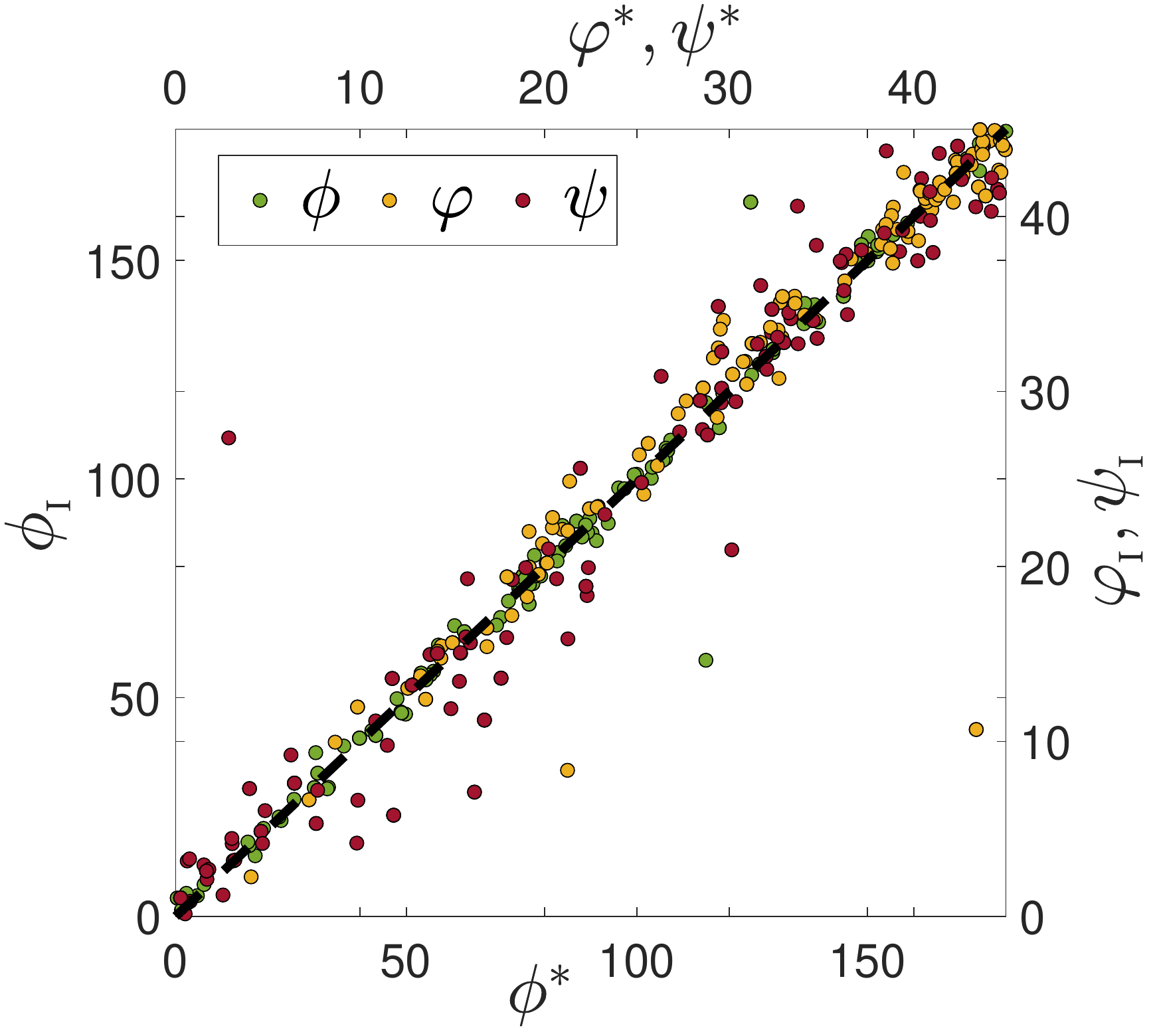}
\caption{The continuum quantities $\phi_{\mbox{\tiny I}}$ (left $y$-axis) and $(\varphi_{\mbox{\tiny I}}, \psi_{\mbox{\tiny I}})$ (right $y$-axis), see Subsect.~\ref{subsec:3D} and Fig.~\ref{fig:fig5} for details, vs.~$\phi^*$ (lower $x$-axis) and $(\varphi^*, \psi^*$) (upper $x$-axis), see Subsect~\ref{subsec:microscopic} for details. The black dashed line corresponds to perfect agreement between the two approaches.}
\label{fig:fig4}
\end{figure}
In Fig.~\ref{fig:fig4} we present the corresponding comparison for many nonlinear quasilocalized modes in 3D (see Appendix~\ref{ap:nlmodes}). In order to distinguish the two approaches, we use the notation $(\phi^*, \varphi^*, \psi^*)$ to refer to the orientation extracted from the microscopic measure defined in this subsection and $(\phi_{\mbox{\tiny I}}, \varphi_{\mbox{\tiny I}}, \psi_{\mbox{\tiny I}})$ for the corresponding quantities extracted from the continuum measure, as done in Fig.~\ref{fig:fig3}a. The comparison in Fig.~\ref{fig:fig4} reveals good agreement between the core orientation extracted from the two approaches, further substantiating both. Next, we directly test the validity of the continuum approach, which allows to extract the relative magnitudes of the core strain components, in addition to the core orientation.
\begin{figure*}[ht]
\centering
\includegraphics[width=0.85\textwidth]{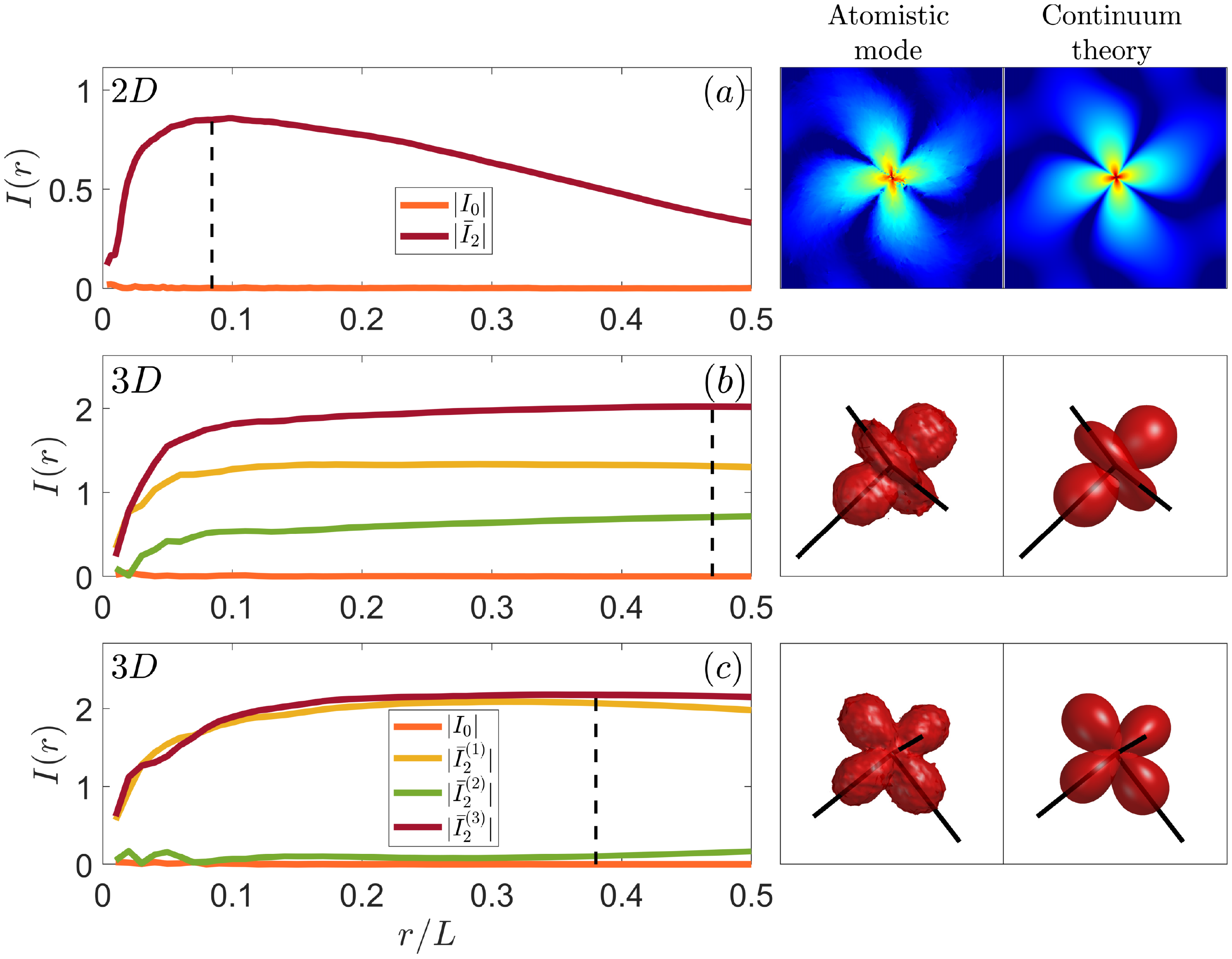}
\caption{(a) The contour integrals of Eqs.~\eqref{eq:I_0}-\eqref{eq:I_2S} for a 2D mode (the same one previously shown in Fig.~\ref{fig:fig3}a) are shown on the left panel. On the right panel, the amplitude of the atomistic mode is shown under ``Atomistic mode'' and mode obtained by the extracted core properties is shown under ``Continuum theory'' (see text for additional details). The core properties are extracted at a distance from the core's center where the largest contour integral (here $\bar{I}_2(r)$) attains its maximum, which is marked by a vertical dashed line. The R-squared correlation coefficient of the atomistic and continuum fields, for $r\!>\!0.15L$, is $R^2\!=\!0.98$. (b)-(c) The contour integrals of Eqs.~\eqref{eq:3DI_0}-\eqref{eq:3DI_m} and Eq.~\eqref{eq:matrix} for two 3D modes are shown on the left panels. The vertical dashed lines, as in (a), mark the distance from the core's center where the largest contour integral attains its maximum. The corresponding R-squared correlation coefficients for $r\!>\!0.15L$ are $R^2\!=\!0.92$ for (b) and $R^2\!=\!0.93$ for (c). On the right panel, surfaces of constant value of the magnitude of each field are shown. The principal directions extracted from the analysis are superimposed (they are identical in both the ``Atomistic mode'' and the ``Continuum theory'' columns). In the continuum theory modes, $2^8$ Fourier modes per spatial dimension have been used in all panels.}
\label{fig:fig5}
\end{figure*}

\subsection{Direct comparison of the continuum theory to atomistic quasilocalized modes}
\label{subsec:full-field}

The continuum approach in 2D and 3D yields the field $\bm{u}^{\circ}(\bm{r})$, i.e.~the one obtained using $\bm{G}^{\circ}(\bm{r})\=\sum_{\bm{q}\in\bm{Q}}e^{2 \pi i \bm{q} \cdot \bm{r}}\,\bm{G}(\bm{q})$ together with Eqs.~\eqref{eq:Green_FT}-\eqref{eq:disp_lr}. The core properties encapsulated in $\bm{\mathcal{E}}^*$ in Eq.~\eqref{eq:disp_lr} are extracted from the atomistic modes obtained in the computer simulations using the continuum theory according to Eqs.~\eqref{eq:I_0}-\eqref{eq:I_2S} in 2D and Eqs.~\eqref{eq:3DI_0}-\eqref{eq:3DI_m} in 3D. Our goal here is to directly compare $\bm{u}^{\circ}(\bm{r})$ to its corresponding atomistic mode $\bm{u}(\bm{r})$, which is a discrete field defined at the particle positions. Since the continuum approach cannot separate the eigenstrains from the $\dbar$-dimensional core volume, i.e.~they are determined up to an over multiplicative factor and only their ratios are accessible, and as in any case $\bm{u}(\bm{r})$ is a normalized field, we normalized hereafter $\bm{u}^{\circ}(\bm{r})$ as well. We then compare the two in a parameter-free manner.

The core properties extraction procedure and the comparison to the atomistic modes are demonstrated in detail in Fig.~\ref{fig:fig5}, in both 2D and 3D. In Fig.~\ref{fig:fig5}a, we consider a 2D mode, and present (left panel) $|I_0(r)|$ and $|\bar{I}_2(r)|$ of Eqs.~\eqref{eq:I_0}-\eqref{eq:I_2S}. As explained in Sect.~\ref{sec:image}, the core properties are extracted at a distance from the core's center where the largest contour integral (here $\bar{I}_2(r)$) attains its maximum. This distance is marked by a vertical dashed line. The extracted core properties are then used to compare $\bm{u}^{\circ}(\bm{r})$ to $\bm{u}(\bm{r})$, where the latter is shown under ``Atomistic mode'' and the former under ``Continuum theory'' on the right panel of Fig.~\ref{fig:fig5}a. Beyond the striking visual resemblance of the two fields, we calculated the R-squared correlation coefficient of the two fields for $r\!>\!0.15L$ (recall that the continuum theory is valid away from the core), yielding $R^2\!=\!0.98$.

In Figs.~\ref{fig:fig5}b-c, we present the analysis of two 3D modes. In this case, $|I_0(r)|$ and $|\bar{I}_{2}^{{\scriptscriptstyle(i)}}(r)|$ (with $i\=1\!-\!3$), cf.~Eqs.~\eqref{eq:3DI_0}-\eqref{eq:3DI_m} and Eq.~\eqref{eq:matrix}, are presented (recall that $\bar{I}_{2}^{{\scriptscriptstyle(3)}}\=-(\bar{I}_{2}^{{\scriptscriptstyle(1)}}+\bar{I}_{2}^{{\scriptscriptstyle(2)}}))$. In these two cases, the contour integrals feature extended plateaus, as predicted by the infinite medium theory, indicating reduced image interaction effects compared to 2D. This is most likely related to the stronger spatial decay of the long-range fields of quasilocalized modes with increasing dimensionality $\dbar$, in line with the findings of~\cite{Rodney2016}. Note that the maximum of the largest contour integral, where the core properties are extracted (indicated by the vertical dashed lines), occurs far from the core itself. This is a direct demonstration of the basic idea developed in this paper, i.e.~that the core properties of quasilocalized modes can be extracted from their far-field behavior.

The quality of the extracted core properties is again quantified by calculating the R-squared correlation coefficient of the atomistic and continuum fields for $r\!>\!0.15L$, yielding $R^2\!=\!0.92$ (Fig.~\ref{fig:fig5}b) and $R^2\!=\!0.93$ (Fig.~\ref{fig:fig5}c). Visual comparisons of the atomistic modes and their continuum counterparts are presented on the right panels of Fig.~\ref{fig:fig5}b-c, where surfaces of constant value of the magnitude of each field are shown. The principal directions extracted from the analysis are superimposed (they are identical in both the ``Atomistic mode'' and the ``Continuum theory'' columns). The strong visual resemblance of the constant value surfaces is in line with the large R-squared correlation coefficients. Finally, while our focus here is on the method for extracting the core properties and not on the physics of the quasilocalized modes themselves, we note that the two modes shown in Figs.~\ref{fig:fig5}b-c reveal quite distinct geometries; while the mode shown Fig.~\ref{fig:fig5}b features a 3D structure characterized by $3$ comparable deviatoric strain amplitudes, the one shown in Fig.~\ref{fig:fig5}c is predominantly planar, where one deviatoric strain amplitude is negligible compared to the other two.

\section{Concluding remarks}
\label{sec:summary}

In this paper we developed an approach for extracting the short-range core properties of quasilocalized modes in glasses, making use of their long-range, power-law elastic fields. In particular, we constructed a set of contour integrals performed on the long-range continuum fields that give access to the short-range core properties. We demonstrated that the long-range fields may experience rotation and suppression due to the periodic boundary conditions commonly employed in computer glass simulations, especially in 2D, and that for computer glasses of typical sizes used in current studies, these finite-size boundary effects may complicate the extraction of the core properties. We subsequently developed a continuum theory of image interactions mediated by the box shape and the periodic boundary conditions, which quantitatively predicted the observed effects on the long-range fields, and allowed the extraction of the core properties. The resulting framework has been tested and validated against a large set of quasilocalized modes in atomistic computer glasses in both 2D and 3D.

The short-range core properties of quasilocalized modes play important roles in the physics of glasses, for example in dissipative plastic deformation, where the quasilocalized modes take the form of irreversible rearrangements. The present paper is methodological in nature, aiming at developing and substantiating a tool that allows the extraction of the core properties in computer glasses. We stress that even though the approach developed in this paper has been tested here on nonlinear quasilocalized modes, it can be equally applied to other quasilocalized modes in glasses. Future studies are expected to use this rather general tool to gain insight into the physics embodied in the core properties, for example their dependence on glassy disorder, their statistical distributions and more. Such studies will also need to face related challenges, such as how to isolate quasilocalized modes in various physical situations (e.g.~during externally driven plastic deformation, where various quasilocalized modes interact).

\acknowledgements

E.B.~acknowledges support from the Minerva Foundation with funding from the Federal German Ministry for Education and Research, the Ben May Center for Chemical Theory and Computation, and the Harold Perlman Family.~E.L.~acknowledges support from the NWO (Vidi grant no.~680-47-554/3259).

\appendix

\section{Inverse Power Law computer glasses}
\label{ap:models}

In this work we have used a 50:50 binary mixture of `large' and `small' particles of mass $m$ that interact via a purely repulsive, inverse power law (IPL) $\sim\! r^{-10}$  pairwise potential. Specifically, the interaction potential $\varphi$ reads
\begin{equation}\label{eq:IPL}
  \varphi\left(r_{ij}\right) = \begin{cases}
               \varepsilon\!\left[\left(\frac{\lambda_{ij}}{r_{ij}}\right)^{10} \!+  \sum\limits_{\ell=0}^{3} c_{2\ell}\left(\frac{r_{ij}}{\lambda_{ij}}\right)^{2\ell}\right], & \frac{r_{ij}}{\lambda_{ij}}<x_c \\
               \qquad \qquad \quad 0 , & \frac{r_{ij}}{\lambda_{ij}} \ge x_c
            \end{cases}
\end{equation}
where $r_{ij}$ is the distance between the $i^{\text{th}}$ and $j^{\text{th}}$ particles, $\varepsilon$ is a microscopic energy scale, $x_c\!=\!1.48$ is the dimensionless distance for which $\varphi$ vanishes continuously up to $3$ derivatives, and the coefficients $c_{2\ell}$ that ensure the said continuity, can be found in Table.~\ref{tab:coeff}. The length parameters are $\lambda^{\mbox{\tiny small}}_{\mbox{\tiny small}}\!=\!\lambda$, $\lambda^{\mbox{\tiny small}}_{\mbox{\tiny large}}\!=\!\lambda^{\mbox{\tiny large}}_{\mbox{\tiny small}}\!=\!1.18\lambda$, and $\lambda^{\mbox{\tiny large}}_{\mbox{\tiny large}}\!=\!1.4\lambda$, where $\lambda$ denotes the microscopic units of length and the subscripts/superscripts correspond to small and large particles (e.g.~$\lambda^{\mbox{\tiny small}}_{\mbox{\tiny large}}\!=\!\lambda^{\mbox{\tiny large}}_{\mbox{\tiny small}}$ corresponds to the length parameter of the interaction between `small' and `large' particles). Times are expressed in terms of $t_0\!=\!\sqrt{m\lambda^2/\varepsilon}$. The densities employed are $\rho\!=\!m N/V\!=\!0.86$ in 2D, and $\rho\!=\!0.82$ in 3D.

\begin{table}[!h]
\caption{\label{tab:coeff}
IPL potential coefficients.}
\begin{ruledtabular}
\begin{tabular}{ccc}
$c_{\mbox{\tiny $0$}}$ & -1.1106337662511798 \\
$c_{\mbox{\tiny $2$}}$ & 1.2676152372297065 \\
$c_{\mbox{\tiny $4$}}$ & -0.4960406072849212 \\
$c_{\mbox{\tiny $6$}}$ & 0.0660511826415732\\
\hline
\end{tabular}
\end{ruledtabular}
\end{table}

We prepared ensembles of computer glasses of this model by first equilibrating the system at high temperature liquid states, followed by performing a continuous quench at rate $\dot{T}\!=\!10^{-3}\varepsilon/k_Bt_0$ from those high temperature liquid states to down below the glass transition temperature $T_g \!\approx\! 0.5 \varepsilon/k_B$, as described in~\cite{cge_paper}. In 2D we have used $N\!=\!102400$, and obtained $1000$ different glassy samples. In 3D we chose $N\!=\!10^6$, obtaining $99$ samples.

\section{Obtaining nonlinear modes}
\label{ap:nlmodes}

The micromechanical objects on which the analysis described in this work was performed, are \emph{nonlinear quasilocalized modes}. These modes were obtained following~\cite{SciPost2016}, by minimizing a cost function $b(\zv)$, that reads
\begin{equation}\label{eq:barrier}
  b(\zv) \equiv \frac{\big(\frac{\partial^2U}{\partial\xv\partial\xv}:\zv \zv\big)^3}{\big(\frac{\partial^3U}{\partial\xv\partial\xv\partial\xv}\cdddot\zv\zv\zv\big)^2} \ ,
\end{equation}
with respect to the putative displacement $\zv$ about the mechanical equilibrium configuration (minimum of $U(\xv)$). Nonlinear quasilocalized modes $\bm{\pi}$ are defined as the displacements $\zv$ for which $b(\zv)$ attains a local minimum, meaning that $\partial b/\partial \zv\big|_{\zv=\bm{\pi}}\!=\!{\bm 0}$. Each such local minimum corresponds to a single nonlinear quasilocalized mode.

To obtain several different nonlinear modes from each of the glassy samples, we have initiated the minimization of $b(\zv)$ with different initial conditions $\zv_\phi$, corresponding to the linear force response to an imposed shear deformation at an angle $\phi$, namely
\begin{equation}\label{initial_conditions}
\zv_\phi = \bigg(\frac{\partial^2U}{\partial\xv\partial\xv}\bigg)^{-1}\cdot \frac{\partial^2U}{\partial\xv\partial\gamma_\phi}\,,
\end{equation}
where $\gamma_\phi$ is a shear strain applied at angle $\phi$ with respect to the Cartesian axes.

In 2D we have used four different biasing angles $\phi\!=\!0^\circ,22.5^\circ, 45^\circ$ and $67.2^\circ$ for each glassy sample, resulting in a total of $\approx\!4$K distinct quasilocalized modes. Some minimizations end with the same quasilocalized mode; we considered only distinct modes in our analyses. The initial conditions in 3D were generated via Eq.~(\ref{initial_conditions}), but this time $\phi$ is understood to represent simple and pure shear in the $x$-$y$, $x$-$z$, and $y$-$z$ planes, resulting in $\approx\!500$ distinct quasilocalized modes.


%

\end{document}